\documentclass[article,preprint,superscriptaddress]{revtex4-1}
\usepackage{graphicx}
\usepackage{dcolumn}
\usepackage{amssymb}
\usepackage{amsmath}
\usepackage{bm}
\usepackage{pifont}
\usepackage{epsfig}
\usepackage{psfrag}
\usepackage[usenames]{color}
\begin{document}

\title{Growing Dynamical Facilitation on Approaching the Random Pinning Colloidal Glass Transition}
\author{Shreyas Gokhale$^{\ast}$}
\affiliation{Department of Physics, Indian Institute of Science, Bangalore - 560012, INDIA}
\author{K. Hima Nagamanasa$^{\ast}$}
\affiliation{Chemistry and Physics of Materials Unit, Jawaharlal Nehru Centre for Advanced Scientific Research, Jakkur, Bangalore - 560064, INDIA}
\author{Rajesh Ganapathy}
\affiliation{International Centre for Materials Science, Jawaharlal Nehru Centre for Advanced Scientific Research, Jakkur, Bangalore - 560064, INDIA}
\author{A. K. Sood}
\affiliation{Department of Physics, Indian Institute of Science, Bangalore - 560012, INDIA}
\affiliation{International Centre for Materials Science, Jawaharlal Nehru Centre for Advanced Scientific Research, Jakkur, Bangalore - 560064, INDIA}
\date{\today}
\draft
\maketitle
\renewcommand{\thefootnote}

Despite decades of research, it remains to be established whether the transformation of a liquid into a glass is fundamentally thermodynamic or dynamic in origin. While observations of growing length scales \cite{weeks2000three,kegel2000direct} are consistent with thermodynamic perspectives like the Random First-Order Transition theory (RFOT) \cite{kirkpatrick1989scaling,lubchenko2007theory}, the purely dynamic approach of the Dynamical Facilitation (DF) theory \cite{hedges2009dynamic,chandler2009dynamics} lacks experimental validation \cite{berthier2011theoretical,biroli2013perspective}. Further, for glass transitions induced by randomly freezing a subset of particles in the liquid phase, simulations \cite{kob2013probing} support the predictions of RFOT \cite{cammarota2012ideal}, whereas the DF theory remains unexplored. Here, using video microscopy and holographic optical tweezers, we show that dynamical facilitation in a colloidal glass-forming liquid unambiguously grows with density as well as the fraction of pinned particles. In addition, we show that heterogeneous dynamics in the form of string-like cooperative motion, which is believed to be consistent with RFOT \cite{starr2013relationship}, emerges naturally within the framework of facilitation \cite{keys2011excitations}. Most importantly, our findings demonstrate that a purely dynamic origin of the glass transition cannot be ruled out.

In spite of significant theoretical \cite{berthier2011theoretical} and experimental \cite{ediger2000spatially,hunter2012physics} advances since the pioneering work of Adam and Gibbs \citep{adam1965temperature}, it is as yet unclear whether the glass transition is thermodynamic or dynamic in nature. This is predominantly because supercooled liquids fall out of equilibrium well before the putative phase transition, be it dynamic or thermodynamic, is reached. This limitation has spurred the search for alternate ways of approaching the glass transition \cite{hedges2009dynamic,cammarota2012ideal,franz1997phase}, of which random pinning has attracted widespread interest in the last two years. Random pinning has mainly been studied within the framework of RFOT, which predicts an ideal glass transition at a finite fraction of pinned particles, for temperatures above the Kauzmann temperature of the unconstrained liquid \cite{cammarota2012ideal}. By comparison, the contrasting perspective of dynamical facilitation has received far less attention. Recently, it was shown that kinetically constrained spin models, which form the conceptual basis of the DF theory, do not exhibit a thermodynamic phase transition with random pinning \cite{jack2012random}. Nonetheless, a systematic analysis of facilitation in the context of real glass-formers is yet to be undertaken, and is imperative to gain a deeper understanding of the random pinning glass transition. 

Since the DF theory has not been validated in experiments even in the absence of pinning \citep{berthier2011theoretical,biroli2013perspective}, we first investigated dynamical facilitation for the unpinned case, in a binary colloidal glass-forming liquid composed of silica particles (see Materials and Methods). According to the DF theory, structural relaxation is dictated by the concerted motion of localized mobile defects, termed excitations, and the slowdown of dynamics is attributed to a reduction in the concentration of these excitations. To identify and characterize excitations, we employed the protocol developed in \cite{keys2011excitations}. A particle was said to be associated with an excitation of size $a$ and `instanton' time $\Delta t$, if it underwent a displacement of magnitude $a$ over a time interval $\Delta t$, and persisted in its initial and final positions for at least $\Delta t$. In practice, excitations were identified by defining for the coarse-grained trajectory $\bar{\textbf{r}}_{i}(t)$ of every particle $i$, the functional
\begin{equation}
h_{i}(t,t_{a};a) = \prod\limits_{t' = t_{a}/2 - \Delta t}^{t_{a}/2} \theta(|\bar{\textbf{r}}_{i}(t+t')-\bar{\textbf{r}}_{i}(t-t')| - a)
\end{equation}
Here, $\theta(x)$ is the Heaviside step function and $t_{a}$, known as the commitment time, is typically chosen to be $\sim$ 3-4 times the mean value of $\Delta t$ for an excitation of size $a$ (See Supplementary Text. Also see Inset to Fig. \ref{Fig1}A). To obtain sufficient statistics, we restricted our analysis to a = 0.5$\sigma_{S}$. We observe that the distributions of instanton times P$_{a}$($\Delta t$) for different area fractions $\phi$ nearly collapse onto the same curve and the mean instanton time $\langle \Delta t \rangle$ remains almost constant (Fig. \ref{Fig1}A). Also, at large $\phi$, $\langle \Delta t \rangle$ is much smaller than the structural relaxation time $\tau_{\alpha}$ (Supplementary Fig. S1), which clearly shows that excitations are localized in time, in accordance with \cite{keys2011excitations}.

To measure the spatial extent of excitations, we first computed the function
\begin{multline}
\mu(r,t,t';a) = \frac{1}{\langle h_{1}(0,t_{a};a) \rangle} \Bigg\langle h_{1}(0,t_{a};a) \sum\limits_{i \neq 1}^{N}|\bar{\textbf{r}}_{i}(t')-\bar{\textbf{r}}_{i}(t)|\delta(\bar{\textbf{r}}_{i}(t) - \bar{\textbf{r}}_{1}(t) - \bar{\textbf{r}}) \Bigg\rangle
\end{multline}   
For $t = -t_{a}/2$ and $t' = t_{a}/2$, $\mu(r,t,t';a)$ yields the displacement density at a distance $r$ from an excitation of size $a$ located at the origin at time $t = 0$, over a time interval $t_{a}$ centred on $0$. From $\mu(r,t,t';a)$, the spatial extent of excitations was then extracted by defining the function
\begin{equation}
F(r;a) = \frac{\mu(r,-t_{a}/2,t_{a}/2;a)}{g(r)\mu_{\infty}(t_{a})} - 1
\end{equation} 
where $g(r)$ is the radial pair-correlation function and $\mu_{\infty}(t_{a}) = \langle|\bar{\textbf{r}}_{i}(t+t_{a}) - \bar{\textbf{r}}_{i}(t)|\rangle$. $F(r;a)$ decays within 8 particle diameters irrespective of $\phi$ (Fig. \ref{Fig1}B), confirming that excitations are spatially localized objects that do not grow on approaching the glass transition \cite{keys2011excitations}. 

A visual indicator of the concentration of excitations $c_{a}$ is the particles' displacement field computed over $\langle \Delta t \rangle = 12$s (Fig. \ref{Fig1}C-E). The decreasing fraction of particles with displacements greater than $a$ (maroon spheres) observed in Figure \ref{Fig1}C-E strongly indicates a lowering in $c_{a}$ with increasing $\phi$. More quantitatively, the concentration of excitations, formally defined as
\begin{equation}
c_{a} = \Bigg\langle \frac{1}{Vt_{a}} \sum\limits_{i=1}^{N} h_{i}(0,t_{a};a)\Bigg\rangle
\end{equation}  
where $V$ is the volume and $N$ is the total number of particles, indeed decreases with $\phi$, in accordance with the DF theory (Fig. \ref{Fig1}F).

The second major ingredient of the DF theory is facilitation, which assumes that new excitations primarily occur in the vicinity of existing excitations. To quantify the degree of facilitation in our system, we computed the facilitation volume 
\begin{equation}
v_{F}(t) = \int\Bigg[\frac{\mu(r,t_{a}/2,t;a)}{g(r)\mu_{\infty}(t-t_{a}/2)} - 1\Bigg]d\textbf{r}
\end{equation} 
We observe that for all $\phi$s considered, $v_{F}(t)$ initially increases with time, reaches a maximum value $v_{F}^{max}$ at time $t_{max}$ and then decreases at longer times (Fig. \ref{Fig1}G). Moreover, both $t_{max}$ and $v_{F}^{max}$ increase with $\phi$ (Fig. \ref{Fig1}G), in complete agreement with recent simulations of atomistic glass-formers as well as spin models \cite{keys2011excitations,elmatad2012manifestations}. These results from colloid experiments provide unequivocal evidence for growing dynamical facilitation on approaching the glass transition, thereby validating the DF theory (See Supplementary Video S1 for a visualization of facilitated dynamics). 

Consistent with RFOT and the Adam-Gibbs theory, a considerable body of evidence supporting the existence of growing dynamic length scales on approaching the glass transition \cite{weeks2000three,kegel2000direct,donati1998stringlike,berthier2005direct,kob2012non} has accumulated over the last two decades. It is natural to wonder if these growing dynamic correlations are influenced by dynamical facilitation. To investigate this possibility, we explored the connection between localized excitations and string-like cooperative motion of mobile particles \cite{donati1998stringlike,keys2007measurement}. We constructed strings from the top 10$\%$ most mobile particles using the procedure described in \cite{donati1998stringlike}. Figure \ref{Fig2}A shows a representative string composed of 13 particles for $\phi = 0.79$, for which $\Delta\tau = t^{\ast} = 66$s, where $t^{\ast}$ corresponds to the maximum of the non-Gaussian parameter $\alpha_{2}(t)$ \cite{weeks2000three}. From their trajectories over $t^{\ast}$, the particles constituting the string can be divided into three independent groups, termed `microstrings' \cite{gebremichael2004particle}, such that motion within each microstring is coherent (Fig. \ref{Fig2}B-D). We observe that in one of the microstrings (Fig. \ref{Fig2}D), each particle $i$ is associated with an excitation for some time $\Delta t_{i} < t^{\ast}$ (Fig. \ref{Fig2}E), which strongly suggests that over $t^{\ast}$, the motion in the other microstrings is facilitated by these excitations. Moreover, the maximum string length $n_{s}^{max}$ increases with the mean separation between excitations (Supplementary Fig. S2). Collectively, these findings show that string-like cooperative motion at longer times emerges in a hierarchical manner, from the short-time dynamics of excitations \cite{keys2011excitations}.

In the presence of random pinning, numerical evidence based on static quantities supports the RFOT scenario \cite{kob2013probing}, whereas that based on dynamic ones does not \cite{jack2013dynamical}. However, the rapid growth of relaxation times with increasing fraction of pinned particles is undisputed \cite{cammarota2012ideal,kob2013probing,jack2013dynamical}. It is therefore natural to ask whether dynamical facilitation accounts for the observed slowdown of dynamics. Indeed, since the DF theory is a parallel approach to RFOT, a systematic investigation of facilitation in the context of random pinning is highly desirable \cite{cammarota2012ideal} and is likely to enrich our understanding of the glass transition. We therefore explored the impact of dynamical facilitation on structural relaxation in the presence of random pinning. Towards this end, we imaged a binary mixture of polystyrene particles at a fixed $\phi \approx 0.71$, which corresponds to the mildly supercooled regime, and used holographic optical tweezers to pin a subset of particles in the field of view (Fig. \ref{Fig3}A. Also see Materials and Methods and Supplementary Video S2). The fraction of pinned particles $f_{p}$ was varied from 0.03 to 0.12. Although these values of $f_{p}$ are too small to induce a glass transition at this $\phi$, they do have a significant impact on the dynamics of the remaining free particles. We see that the structural relaxation time $\tau_{\alpha}$ (Supplementary Fig. S3), increases by a factor of 4 relative to its value in the unpinned case (Fig. \ref{Fig3}B). This slowing down of dynamics is consistent with recent simulations \cite{kob2013probing,jack2013dynamical}. To verify whether this slowdown is a consequence of growing facilitation, we repeated the analysis shown in Figure \ref{Fig1} for the pinned case. We observe that the concentration of excitations $c_{a}$ decreases by more than a factor of 3 with $f_{p}$ (Fig. \ref{Fig3}B), whereas the facilitation volume $v_{F}(t)$ shows a concomitant increase (Fig. \ref{Fig3}C), in close analogy to their behavior with increasing $\phi$ (Fig. \ref{Fig1}F-G). While the assumptions of the DF theory are expected to be valid close to the glass transition \citep{berthier2011theoretical}, our results show that even in the mildly supercooled regime, $c_{a}$ as well as $v_{F}(t)$ are strongly correlated with the structural relaxation time.  In striking resemblance to the unpinned case, our findings place the dynamical facilitation theory on a firm experimental footing even in the context of the random pinning glass transition. 

Next, we examined the interplay between facilitation and string-like cooperative motion in the presence of random pinning. We observe that the behavior of string-like cooperative motion as a function of $f_{p}$ is remarkably similar to its variation with $\phi$. In particular, the distribution of string lengths $P(n)$ is nearly exponential (Fig. \ref{Fig3}D), and the average string length $n_{s} = \Sigma n^2P(n)/\Sigma nP(n)$ exhibits a maximum as a function of time (Fig. \ref{Fig3}E. See Supplementary Fig. S2 for corresponding results in the unpinned case). Further, the time at which $n_{s}$ reaches a maximum, $\Delta\tau_{max}$, increase systematically with $f_{p}$ (Fig. \ref{Fig3}E). The maximal string length  $n_{s}^{max}$ at large $f_{p}$ is greater than its value for the unconstrained liquid (Fig. \ref{Fig3}E). However, for the $\phi$ and the range of $f_{p}$ considered here, it is not possible to conclude definitively, whether the string length grows on approaching the random pinning glass transition. Experiments and simulations at higher values of $\phi$ and $f_{p}$ are necessary in order to clarify this scenario. 

Our experiments have revealed the increasing importance of dynamical facilitation on approaching the random pinning glass transition. The obvious consequence of our findings is that one cannot \textit{a priori} rule out the existence of a purely dynamic `space-time' phase transition from an active liquid to an inactive glassy state \cite{hedges2009dynamic}, even in the context of random pinning. Moreover, even if a thermodynamic phase transition as a function of $f_p$ exists, it is possible for dynamical facilitation to be the dominant mechanism of structural relaxation. In this context, it would be especially interesting to explore the connection between facilitation and growing \textit{static} point-to-set correlations \citep{biroli2008thermodynamic}. It is tempting to speculate that dynamical facilitation may also play a role in the generic decoupling of static and dynamic length scales \cite{kob2012non,charbonneau2013decorrelation}. Further, since $c_{a}$ is known to be an order parameter for space-time phase transitions \cite{speck2012constrained}, the decrease in $c_{a}$ with $f_{p}$ implies that changing $f_{p}$ influences the system's proximity to a dynamic phase transition. Whether or not this dynamic phase transition coincides with the thermodynamic one predicted by RFOT could well be a test for validating both the RFOT and the DF scenarios. We expect a confluence of experimental, theoretical and numerical research aimed at exploring these questions to emerge in the wake of our findings.  

\section*{Materials and Methods}
For experiments without pinning, the system comprised of N$_{L}$ large and N$_{S}$ small silica colloids of diameters $\sigma_{L}$ $\sim$ 970 nm and $\sigma_{S}$ $\sim$ 760 nm respectively. The size ratio $\sigma_{L}$/$\sigma_{S}$ $\sim$ 1.3 and the number ratio N$_{L}$/N$_{S}$ $\sim$ 0.66 effectively suppressed crystallization. Dilute samples were loaded in a wedge-shaped cell \cite{nagamanasa2011confined} and the area fraction $\phi$ was tuned systematically by controlled sedimentation of particles to the monolayer thick region of the wedge. For experiments with pinning, we used a binary mixture of polystyrene spheres of diameters 1.4 $\mu$m and 1.05 $\mu$m with number ratio N$_{L}$/N$_{S}$ $\sim$ 2.3.  The holographic optical tweezers set up consisted of a linearly polarized constant power (800 mW) CW laser (Spectra-Physics, $\lambda$ = 1064 nm). A spatial light modulator (Boulder Nonlinear Systems, Inc.) was used to generate upto 120 traps ($f_{p}$ = 0.12) at randomly chosen locations. Only large polystyrene particles were pinned. The laser was switched on after fixing the trap positions and equilibrating the sample to ensure simultaneous activation of the traps. The average trap strength was sufficient to ensure that the pinned particles did not escape over the duration of the experiment, as evidenced by the difference in mean squared displacements of free and pinned particles (Supplementary Fig. S4). For both experiments, the samples were imaged using a Leica DMI 6000B optical microscope with a 100X objective (Plan apochromat, NA 1.4, oil immersion) and images were captured at 5 fps for 1-3 hrs. The 66 $\mu$m x 49 $\mu$m field of view typically contained $>$ 4000 particles for the silica sample and $\sim$ 1200 particles for the polystyrene sample. Standard algorithms were used for particle tracking \cite{crocker1996methods} and quantities of interest were extracted using codes developed in-house.

\section*{Acknowledgements}
The authors thank Gilles Tarjus, Sriram Ramaswamy, C. Patrick Royall, Francesco Sciortino and Chandan Dasgupta for useful discussions. We thank V. Santhosh for help in setting up the holographic optical tweezers, Mamta Jotkar for synthesizing silica colloids and Chandan Mishra and Amritha Rangarajan for synthesizing polystyrene colloids. S.G. thanks the Council for Scientific and Industrial Research (CSIR), India for a Shyama Prasad Mukherjee Fellowship. K.H.N. thanks CSIR, India for a Senior Research Fellowship. R.G. thanks the International Centre for Materials Science (ICMS) and the Jawaharlal Nehru Centre for Advanced Scientific Research (JNCASR) for financial support and A.K.S. thanks Department of Science and Technology (DST), India for support under J.C. Bose Fellowship. 

\subsection*{Author Contributions}
S.G. and K.H.N contributed equally to this work.\\
$^{\ast}$ Corresponding author
 

\newpage
\begin{figure*}
  \centering
  \includegraphics[height=14cm]{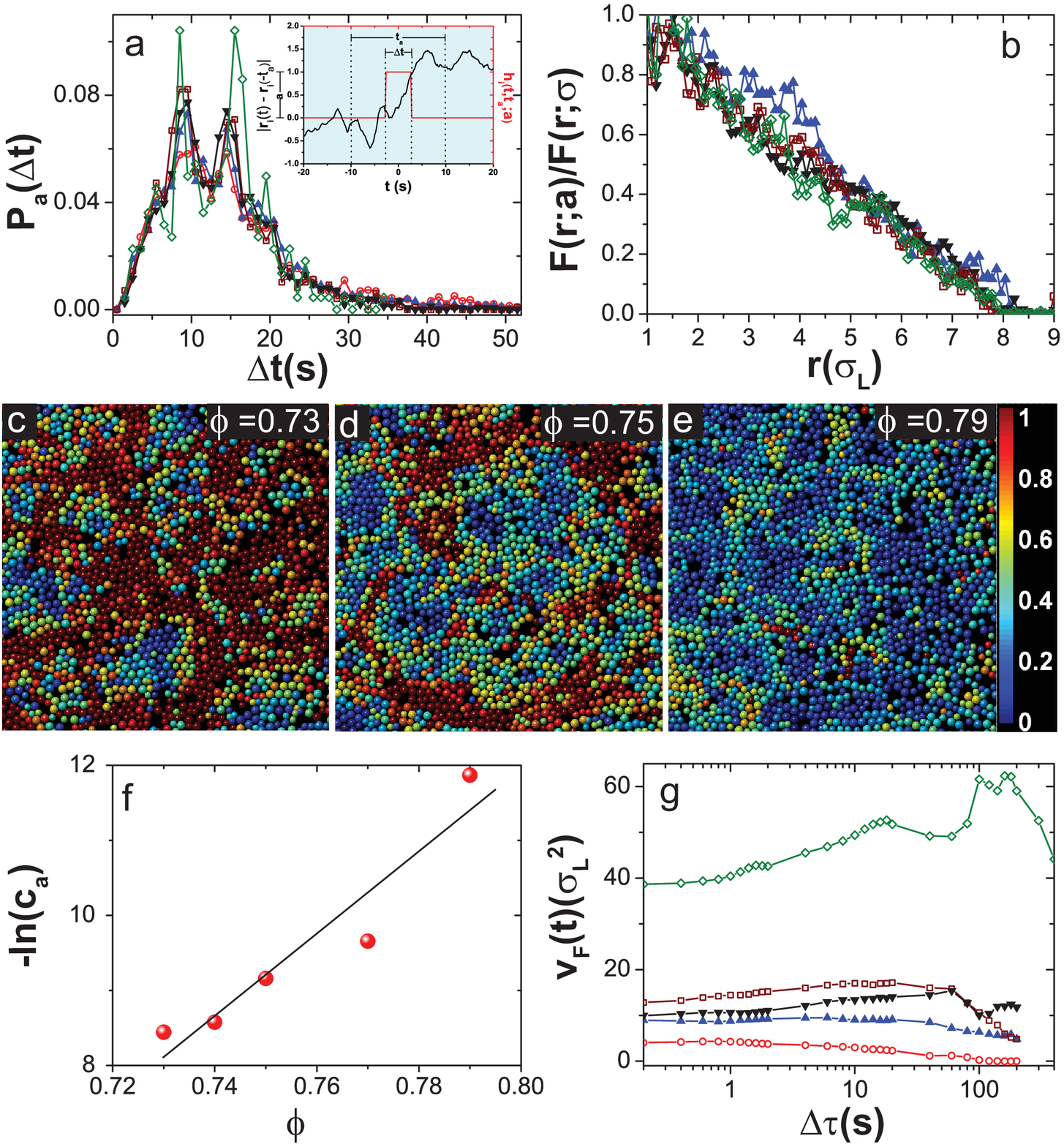}
  \caption{Dynamical Facilitation in the absence of pinning. \textbf{a.} Distribution of instanton times $P_{a}(\Delta t)$ for for $\phi =$ 0.73 ({\color{Red} $\boldsymbol \circ$}), $\phi =$ 0.74 ({\color{Blue} $\boldsymbol \blacktriangle$}), $\phi =$ 0.75 ({\color{Black} $\boldsymbol \blacktriangledown$}), $\phi =$ 0.77 ({\color{Brown} $\boldsymbol \square$}) and $\phi =$ 0.79 ({\color{Green} $\boldsymbol \diamond$}), showing that excitations are localized in time. Inset to \textbf{a.} Representative sub-trajectory of a particle coarse-grained over 2s, shown in black, and the corresponding functional $h_{i}(t,t_{a};a)$ shown in red. The instanton time duration $\Delta t$ and the commitment time $t_{a}$ are marked by dotted lines. The displacement is measured in units of $a = 0.5\sigma_{S}$ and the trajectory has been shifted arbitrarily along the Y-axis to make the rise in displacement coincide with region of non-zero $h_{i}(t,t_{a};a)$. \textbf{b.} The function $F(r;a)$ normalized by its value at $r = \sigma_{L}$ for $\phi =$ 0.74 ({\color{Blue} $\boldsymbol \blacktriangle$}), $\phi =$ 0.75 ({\color{Black} $\boldsymbol \blacktriangledown$}), $\phi =$ 0.77 ({\color{Brown} $\boldsymbol \square$}) and $\phi =$ 0.79 ({\color{Green} $\boldsymbol \diamond$}), showing that excitations are localized in space. \textbf{c-e.} Displacement field of particles for $\phi = $0.73 (\textbf{c}), $\phi =$ 0.75 (\textbf{d}) and $\phi =$ 0.79 (\textbf{e}). Displacements are normalized by the excitation size $a$. Particles coloured in maroon indicate displacements $\geq a$. \textbf{f.} Concentration of excitations $c_{a}$ vs $\phi$. \textbf{g.} Facilitation volume $v_{F}(t)$ for various $\phi$s. The symbols and colors in (\textbf{g}) are same as those in (\textbf{a-b}).}
  \label{Fig1}
\end{figure*}

\newpage
\begin{figure*}
  \centering
  \includegraphics[width=\textwidth]{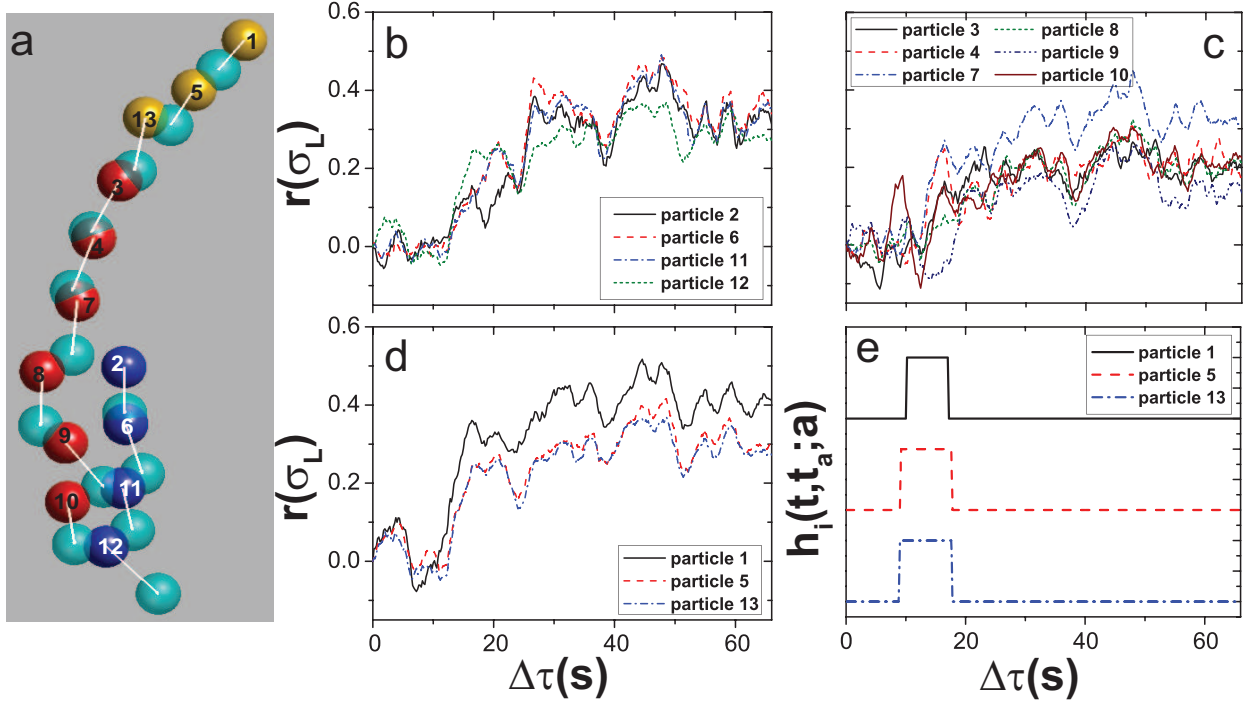}
  \caption{Connection between excitation dynamics and microstrings. \textbf{a.} Schematic of a representative string composed of 13 particles for $\phi = 0.79$ constructed over a time interval $\Delta\tau = t^{\ast}=$ 66s. Particles in yellow, red and blue correspond to initial positions of particles belonging to three independent microstrings. Cyan spheres represent final positions of the particles. \textbf{b-d.} Trajectories of particles corresponding to the blue (\textbf{b}), red (\textbf{c}) and yellow (\textbf{d}) microstrings over time $t^{\ast}$. In (\textbf{b-d}), the particle labels are identical to those in (\textbf{a}). \textbf{e.} The functional $h_{i}(t,t_{a};a)$, where $a = 0.3\sigma_{S}$, corresponding to the trajectories shown in (\textbf{d}) for the yellow microstring in (\textbf{a}).}
  \label{Fig2}
\end{figure*}

\newpage
\begin{figure}
  \centering
  \includegraphics[height=17cm]{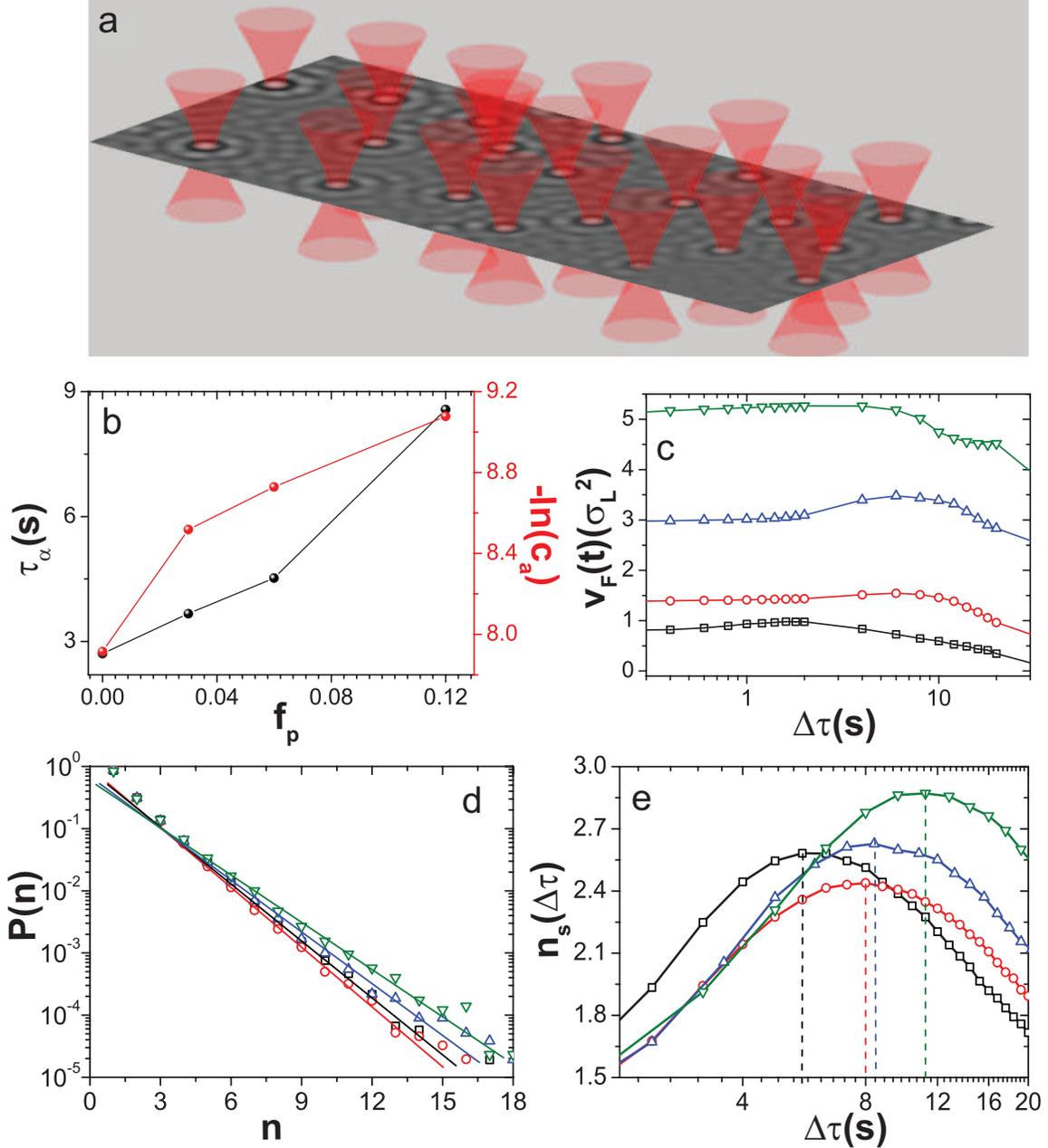}
  \caption{Effect of Random Pinning. \textbf{a.} Schematic of the trapping potentials (shown in red) created by the holographic optical tweezers. The underlying image represents a small portion of the field of view for $f_{p}$ = 0.06. The image has been generated by averaging over $\sim 15\tau_{\alpha}$. Pinned particles appear bright due to their low mobility and high overlap with initial positions. \textbf{b.} Structural relaxation time $\tau_\alpha$ for free particles (black) and concentration of excitations $c_{a}$ (red) as a function of the fraction of pinned particles $f_{p}$. \textbf{c.} Facilitation volume $v_{F}(t)$ for $f_{p} =$ 0 ({\color{Black} $\boldsymbol \square$}), $f_{p} =$ 0.03 ({\color{Red} $\boldsymbol \circ$}), $f_{p} =$ 0.06 ({\color{Blue} $\boldsymbol \triangle$}) and $f_{p} =$ 0.12 ({\color{Green} $\boldsymbol \bigtriangledown$}). \textbf{d.} Distribution of string lengths $P(n)$ for $\Delta\tau$ corresponding to the maximum string length and \textbf{e.} String length $n_{s}(\Delta\tau)$ for various $f_{p}$. The colors and symbols in (\textbf{d}) and (\textbf{e}) are identical to those in (\textbf{c}).}
  \label{Fig3}
\end{figure}
\end{document}